\documentclass[final,3p,times,twocolumn]{style}

\usepackage{graphicx}
\usepackage{amsmath}
\usepackage[english]{babel}
\usepackage{lineno}
\usepackage{url}
\usepackage{hyperref}
\usepackage{bm}
\usepackage{booktabs}

\usepackage{tcolorbox}
\usepackage{dblfloatfix} 
\usepackage{framed}
\usepackage{subcaption}
\usepackage{enumitem}

\setcitestyle{semicolon,aysep={,}}

\begin{document}

\begin{frontmatter}

\title{Multifractal analysis of racially-constrained population patterns and residential segregation in the US cities}

\author[First]{T. F. Stepinski\corref{cor2}}
\ead{stepintz@uc.edu}

\author[Second]{Anna Dmowska}
\ead{dmowska83@gmail.com}

\address[First]{Space Informatics Lab, University of Cincinnati, Cincinnati, OH, USA.}

\address[Second]{Institute of Geoecology and Geoinformation, Adam Mickiewicz University, Poznan, Poland}

\cortext[cor2]{Corresponding Author}

\begin{abstract}
A phenomenon of racial segregation in U.S. cities is a multifaceted area of study. A recent advancement in this field is the development of a methodology that transforms census block-level population count-by-race data into a grid of monoracial cells. This format, conceptually closer to reality, enables novel approaches to assessing residential segregation, particularly the heterogeneity of segregation within a city. This paper leverages such a grid for the detailed quantification of race-constrained population patterns, allowing for the calculation and mapping of residential binary segregation patterns within arbitrarily defined regions. 
A key innovation presented here is the application of Multifractal Analysis (MFA) to quantify the residency patterns of race-constrained populations. The residency pattern is characterized by a multifractal spectrum function, where the independent variable is a local metric of pattern's \textit{gappiness}, and the dependent variable is proportional to the size of the sub-pattern consisting of all locations having the same value of this metric. In the context of binary populations (a race-constrained population and the remaining population), the gappiness of the race-constrained population's pattern is intrinsically linked to its segregation.
This paper provides a comprehensive description of the methodology, illustrated with examples focusing on the residency pattern of Black population in the central region of Washington, DC. Further, the methodology is demonstrated using a sample of residency patterns of Black population in fourteen large U.S. cities. By numerically describing each pattern through a multifractal spectrum, the fourteen patterns are clustered into three distinct categories, each having unique characteristics. Maps of local gappiness and segregation for each city are provided to show the connection between the nature of the multifractal spectrum and the corresponding residency and segregation patterns. This method offers an excellent quantification of race-restricted residency and residential segregation patterns within U.S. cities.
\end{abstract}


\end{frontmatter}


\section{Introduction}
\label{S:1}
Urban populations in the United States (US) exhibit both multiracial diversity and segregation \citep{logan2004, Parisi2011, Lee2014a}. The issue of racial segregation was initially framed in the context of the separation of the Black minority from the White majority \citep{farley1994, massey1993,massey1995}. Consequently, most research on quantitative methods for assessing segregation has been focused on US urban areas (hereafter referred to as cities). In this paper, we work with racially-constrained population patterns -- spatial patterns determined by the locations of residences of inhabitants sharing a common race or ethnicity \citep{Louf2016}. When a specific race is chosen for the analysis, we refer to the pattern as the focus population's pattern. Residential segregation denotes the extent to which different race-constrained populations reside separately in distinct parts of the city \citep{Dadashpoor2024}. Hereafter, we will simply refer to it as segregation.

\begin{figure*}[t]
	\centering
	\includegraphics[width=0.93\textwidth]{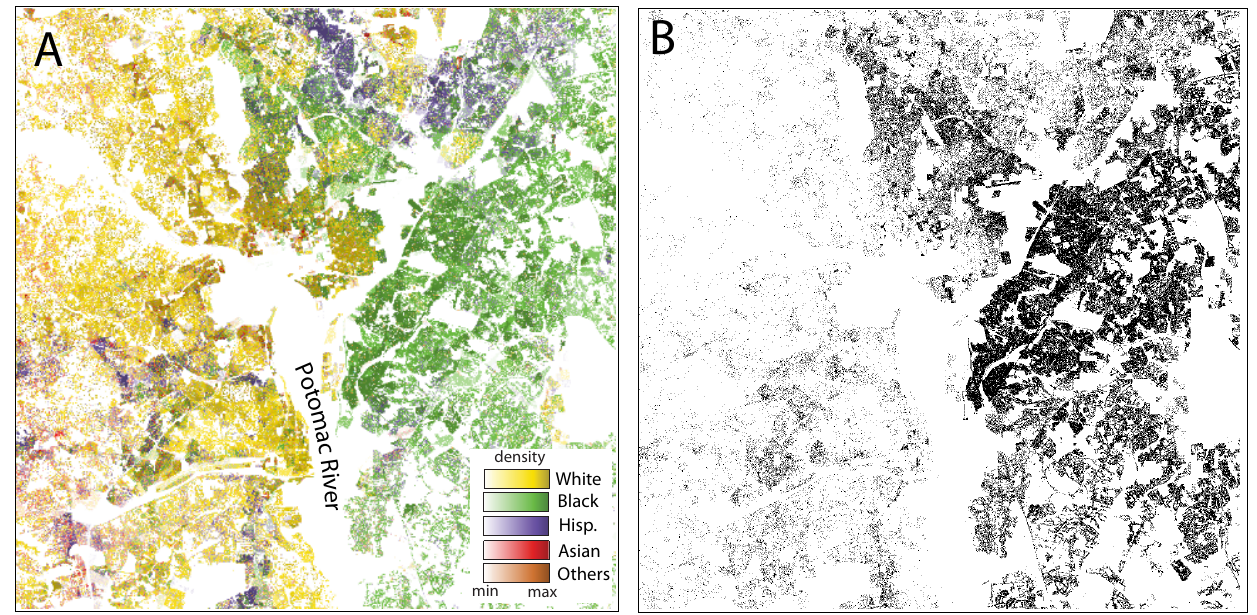}
	\caption{(A) Racial landscape depicting a central region of the Washington, DC; the area shown is $31\times 31$ km with a resolution of 30 meters. The cell's color indicates race, while the shade of the color indicates population density. (B) Spatial distribution of the Black population; black cells correspond to the green cells of all shades in panel (A). All cells inhabited by the Black population are colored black regardless of density. Due to the small cell size, areas of lower density may appear gray due to an optical illusion.  
\label{dc}}
\end{figure*}

Due to privacy concerns, US population data provided by the U.S. Census Bureau is presented as aggregated units of variable sizes and uneven boundaries. This format contrasts with the actual spatial distribution of populations, which may be more accurately conceptualized as a point pattern (representing addresses of residences) rather than as counts within areal units. Researchers thus face a choice: they can either use the aggregated data as it is or employ it to construct models that more closely approximate point patterns.
 
The vast majority of existing research opts for the former approach, utilizing aggregated data directly (for a review of methods based on this data format, see \citep{yao2019spatial}). These methods quantify segregation within an entire metropolitan area or a county by calculating a single metric known as a segregation index. Various segregation indices have been developed, as detailed in \citep{yao2019spatial} and the introduction in \citep{Dmowska2020}. Segregation indices are predominantly used in longitudinal studies to monitor changes in cities' segregation over time \citep{Bellman2018,elbers2021,Farrell2008,Frey2022,Lee2014a}. The primary appeal of segregation indices lies in their ease of use. However, a significant drawback is that they reduce a spatially complex phenomenon to a single-number description, thereby providing no insight into the heterogeneity of segregation within a county or metropolitan area.

One alternative to the direct use of aggregated area data is gridded data. In the US, high-resolution population grids provide densities of various racial populations at a 30-meter resolution \citep{Dmowska2017b}. These grids can be downloaded from \url{https://www.socscape.edu.pl}. However, because the original data source for these grids is the US Census block-level aggregates, the grid cells are multiracial -- they contain vectors of densities for all racial/ethnic populations. Thus, conceptually, these grids are closer to aggregated data than to a spatial point pattern.

Only a small minority of past research has explored modeling the spatial distribution of populations. One method \citep{Masias2024} estimates a 2D probability distribution for each population over a region of interest; this approach does not resemble a spatial point pattern. Another recent method, termed ``racial landscape,'' spatially disaggregates census block's population into a grid of monoracial cells \citep{Dmowska2020,Dmowska2024} (an example of a racial landscape is shown in Fig.~1A). In this method, each cell contains only individuals of the same race. The racial landscape is similar to one-dot-per-person maps \citep{roth2010dot,Dmowska2019}, which depict spatial point patterns with dots labeled by the race of each inhabitant. Both the racial landscape and one-dot-per-person map model the distribution of races within the small extent of a census block, effectively transforming the data into a format more akin to reality.
The gridded nature of the racial landscape offers an advantage over one-dot-per-person maps by providing a structured data format, which facilitates a simpler and faster analysis.

\begin{figure*}[htbp]
    \begin{minipage}{2.1 \columnwidth}
    \begin{framed}
 {\bf Box 1. Terms and definitions}
 
 \vspace{1.2mm}    
\begin{tabular}{p{7.5cm}    p{7.5cm} }
{\it cell} -- areal unit of input data & {\it pattern} -- a set of cells having the same racial ID\\
{\it box} -- areal unit of a measuring grid & $\epsilon$-{\it box} -- a box of size $\epsilon$ \\
{\it empty box}  -- a box covering uninhabited cells & {\it uninhibited box} -- the same as {\it empty box}\\
{\it scale} ($\epsilon$) -- the size of the box & {\it scaling} -- power law relation between a variable and $\epsilon$\\ 
{\it window} -- local area of the size 16$\times$16 cells & {\it neighborhood} -- the same as {\it window} \\
{\it density} - a value in a cell or a box & {\it probability} -- density/total population count\\
$\alpha$ --exponent of a local scaling of density & $\alpha_{\rm D}$ -- $\alpha$ computed from a data in a {\it window}\\
{\it sub-pattern} -- all cells with same racial ID and $\alpha_{\rm D}$ & {\it D} -- fractal dimension  \\ 
 $D_q$ -- generalized fractal dimension of order $q$ &  $f$ -- $D$ of a sub-pattern derived analytically  \\
 $f_{\rm D}$ -- $D$ of a sub-pattern derived computationally & $S_i$ -- local segregation metric  \\
 $D_{\rm fsp}$ -- local fractal dimension of focus pop. & $D_{\rm rp}$ -- local fractal dimension of remaining pop.  \\
 {\it fractal} -- pattern exhibiting a single scaling  &   {\it multifractal} -- pattern exhibiting multiple scalings \\
\end{tabular}
\end{framed}
    \end{minipage}
\end{figure*}

In this paper, we propose applying multifractal analysis (MFA) to racial landscape data to quantify the residency pattern of a focus population and to assess the heterogeneity of segregation of this population from the rest of the population across the region of interest. Figure~\ref{dc}A presents an example of a racial landscape data for a centrally located region of the Washington, DC. In this map cells are monoracial, colors indicate races, and their shades indicate densities. In Figure~\ref{dc}B only cells associated with Black race (green colors of all shades in Figure~\ref{dc}A) are shown. This is a residency pattern of the focus (Black) population. The pattern shows spatial intermittency and clustering on multiple scales -- telltale signs of a multifractal pattern.

Spatial datasets previously analyzed using MFA are built-up area data \citep{Murcio2011,Chen2013,Song2019}, street intersection point pattern (SIPP) data \citep{Murcio2015}, patterns of housing prices \citep{lengyel2023roughness}, and remotely-sensed night-light data \citep{Ozik2005}. MFA has also been used to quantify population patterns within urban areas \citep{Chen2013,ChenFeng2017,tannier2005fractals,lengyel2023roughness} and on regional or larger scales \citep{Appleby1996,Adjali2001,Mannersalo1998,Semecurbe2016,VegaOrozco2015}. However, to our knowledge, it has not been applied to the quantification of constrained populations. Using MFA to quantify race-constrained population not only quantifies this race residency pattern's but also allows for a detailed quantification of local segregation between the focus population and the remaining population.

Our first objective is to quantify the pattern formed by the residency of the focus population using a multifractal spectrum. A multifractal spectrum is a function $f(\alpha)$, where, informally, $\alpha (x,y)$ represents a metric of gappiness of the pattern at location $(x,y)$, and $f(\alpha)$ indicates the size of the sub-pattern exhibiting a given value of gappiness metric (e.g., a size of the subset of the entire pattern where $\alpha (x,y) = 1.5$). Thus, $f(\alpha)$ provides a comprehensive description of the pattern.

However, a multifractal spectrum, as presented in the existing MFA literature \citep{Salat2017,Murcio2015,Chen2013}, cannot be used to construct a useful map of $\alpha (x,y)$  \citep{Theiler1990}. Given the importance of mapping in our analysis, we introduce a discrete multifractal spectrum $f_{\rm D} (\alpha_{\rm D})$, which facilitates mapping the pattern's texture. Quantitative comparisons between patterns formed by the focus population's residency in different cities or within the same city over time can be achieved by comparing $f_{\rm D} (\alpha_{\rm D})$ for the respective patterns.

Our second objective is to quantify local segregation between the focus population and the remaining population. Segregation $S_i$ at location $i$ is quantified by comparing the spatial patterns at the same location $i$ for both the focus population and the remaining population. The quantity $S$, comprising $S_i$ values for $i = 1, \dots, n$ spatial windows covering the entire region of interest, is mappable. A notable feature of $S$ is its ability to distinguish between locations where the focus population is segregated due to the absence of other populations and locations where segregation occurs due to the absence of the focus population itself.

This paper is primarily methodological, focusing on the application of MFA to the study of racial segregation. As such, the second section, which elaborates on the methodology, may be the most valuable. In the third section, we apply our method to a sample of central regions in fourteen large US cities. Finally, we cluster the set of fourteen discrete multifractal spectra into three classes. Each class groups similar spectra and thus regions with patterns having similar statistics of gappiness. Ulltimately, the classes represent types of the spatial organization of segregation heterogeneity. Then we show maps of $\alpha_{\rm D} (x,y)$ and $S_i$ for all fourteen regions, organized by class membership in order to demonstrate visually differences between the three classes of spatial organizations of segregation.

\section{Multifractal analysis}
Box~1 list terms and definitions often used throughout this paper. 

A \textit{fractal} \citep{mandelbrot1967} is a geometric object (in our case, a 2D pattern) that exhibits a single scaling rule repeating at every scale, characterized by a single fractal dimension. An example of a fractal in nature is the leaf structure of a fern. Conversely, a \textit{multifractal} consists of multiple interwoven fractal sets, each governed by its own distinct scaling rule, and characterized by a spectrum of fractal dimensions. Mountain topography serves as a natural example of a multifractal.  Another example of a multifractal is the spatial distribution of Black residency in Washington, DC, shown in Figure~\ref{dc}B.

The standard multifractal analysis (MFA) procedure culminates in obtaining the multifractal spectrum \citep{Harte2001} through the following steps:
\begin{enumerate}
    \item Set up a series of nested grids to measure scaling exponents of density and other variables.
    \item Calculate Renyi’s generalized fractal dimensions.
    \item Calculate multifractal spectrum.
\end{enumerate}

For a comprehensive review of the mathematical procedures required for these steps, refer to Salat et~al. \citep{Salat2017}. Additional valuable references include Murcio et~al. \citep{Murcio2015} and Lopes and Betrouni \citep{Lopes2009}. In this paper, we will not replicate the formalism of the MFA. Instead, all necessary variables and their mathematical formulations are consolidated in Box~2.

In \textbf{Step 1}, a series of progressively coarser grids are utilized to measure density at different scales using the box-counting method \citep{Bouligand1929}. We refer to the unit of a grid as a {\it box}. In the finest grid the size of the box is equal to the size of a cell in the input data.  Progressively coarser grids are generated by aggregating four smaller boxes from the grid at the preceding level of the hierarchy. The grids, their boxes, and values of variables in the boxes are indexed by the size of the box $\epsilon$. The value of $\epsilon$ is set to 1 for the coarsest resolution, consisting of a single box at level $lev = {\rm lev}_{min} = 1$, and decreases according to the power law $\epsilon_{\rm lev} \sim 2^{-({\rm lev}-1)}$, reaching its smallest value of $\epsilon$ at level $lev = {\rm lev}_{max}$.

\begin{figure}[htbp]
    \begin{minipage}{\columnwidth}
    \begin{framed}
 {\bf Box 2. List of major quantities in MFA}
 
 \vspace{0.6mm}    
{\sl Partition function:} This function assesses how different parts of a density distribution contribute to the overall structure at various scales.
\begin{equation}
     Z(q, \epsilon) = \sum_i p_i(\epsilon)^q
\end{equation}  
{\sl Effective number:} Represents the contribution of densities \( p_i \) in a way akin to box-counting, but weighted by the distribution of densities.
\begin{equation}
     N_{\rm eff}(q, \epsilon) = \sum_i p_i (\epsilon)^\frac{1}{1-q} =Z(q, \epsilon)^\frac{1}{1-q}\ \ \ q\neq 1
\end{equation}
{\sl  Renyi's generalized fractal dimensions} \citep{Renyi1961}:  These dimensions describe the scaling behavior of the density across different regimes within the pattern, where each regime is defined by the densest cells in \( N_{\rm eff}(q, \epsilon) \).
\begin{equation}
D_q =
\begin{cases}
-{\displaystyle \lim_{\epsilon\to 0}} \frac{\log N_{\rm eff}(q, \epsilon)}{\log\epsilon}  & \text{if}\ \  q \neq 1 \\
\\
-{\displaystyle \lim_{\epsilon\to 0}} \frac{-\sum_{i=1}^n p_i(\epsilon) \log p_i(\epsilon) }{\log\epsilon}  & \text{if}\ \  q=1 
\end{cases}    
\end{equation} 
{\sl Mass exponent:} The mass exponent \( \tau(q) \) describes the scaling behavior of the partition function; \( Z(q,\epsilon) \sim \epsilon^{\tau(q)} \).
\begin{equation}
    \tau(q) = D_q (q-1) 
\end{equation}
{\sl Local Scaling Exponent:} The function \( \alpha(q) \) represents the local scaling exponent of density, \( p_i(q) \sim \epsilon^{\alpha_i(q)} \). It describes how the scaling behavior of density changes from one location to another.
\begin{equation}
    \alpha(q) = \frac{d D_q}{d q}
\end{equation}
{\sl Local Fractal Dimension:} This function is the fractal dimension of the subset of the pattern where the local scaling exponent \( \alpha \) has a specific value. It indicates how the geometric complexity of the pattern varies with location.
\begin{equation}
    f(q) = q \alpha(q) +\tau (q)
\end{equation}  
\end{framed}
    \end{minipage}
    \label{fig:label}
\end{figure}

Dividing the values of non-empty grid boxes by the total population yields normalized population densities, denoted as $p_i (\epsilon)$. These values represent the probability of finding a specific population density within a cell of the $\epsilon$-grid, and their sum across the grid is equal to 1. The probability distribution function (PDF) at a given scale $\epsilon$ is represented by $\{p_1, \dots, p_n\}_{\epsilon}$, where $p_i$ is both the density and its corresponding probability for cell $i$. The variable $n(\epsilon)$ represents the number of linearly indexed non-empty $\epsilon$-boxes. PDFs vary with $\epsilon$ and must be indexed accordingly, as they differ for each hierarchical level.

Our input data is represented by a square grid with a normalized linear dimension of 1 (scale $\epsilon_1 = 1$). We utilize 10 additional scales, thus ${\rm lev}_{max} = 11$, with the smallest scale being $\epsilon = 1/1024$.

In \textbf{Step 2} we calculate Renyi’s generalized fractal dimensions \citep{Renyi1961}. This involves calculating values of the partition function \citep{Halsey1986} (Eq.~1) for different values of order parameter $q$, calculating effective number \citep{Hill1973} (Eq.~2), and, finally, calculating generalized dimensions $D_q$ themselves (Eq.~3). An effective number $N_{\rm eff}(q, \epsilon)$ is an approximate number of $\epsilon$-boxes that contribute significantly to $Z(q, \epsilon)$. Thus, $q$ can be thought of as a dial for selecting boxes by constraining their density. If $q=0$, $N_{\rm eff}(q, \epsilon)= n(\epsilon)$ -- all non-empty boxes are included. Increasing the value of $q$ results in selecting only the boxes having densities above a threshold, $p_q$ whose value is increasing with $q$. A generalized fractal dimension, $D_q$, is a fractal dimension of a pattern 
restricted to boxes with densities $p_i > p_q$. We compute values of  $D_q$ using the method detailed in Appendix C of \citep{Murcio2015};  we use the following 31 values of $q$, $\{-10, -9, \dots, 0, \dots, 19, 20\}$.

In \textbf{Step 3} we calculate the multifractal spectrum of a pattern. The multifractal spectrum is a parametric function $\{\alpha(q_i), f(q_i)\}$, where $\alpha (q)$ and $f(q)$ are defined by Eqs.~4, 5 and 6. Calculating values of $\alpha(q_i)$ (Eq~5) requires computing a derivative. We first fit a spline to the $\{q_i, D_{q_i}\}$ points using our 31 values of $q$ (see above), and compute the derivative analytically from the spline.

\begin{figure}[t]
    \centering
      \includegraphics[width=8cm]{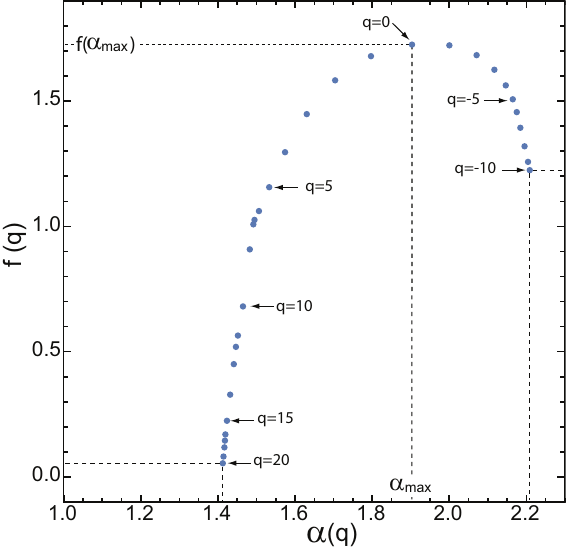}
    \caption{Multifractal spectrum of the Black sub-population pattern in Washington, DC.}
    \label{multifractalSpectrum}
\end{figure} 

Figure~\ref{multifractalSpectrum} shows the multifractal spectrum for the Black residency pattern illustrated in Figure~1B. The blue dots represent the points $\{\alpha(q_i), f(q_i)\}$ for 31 values of $q$.  This spectrum offers a concise quantitative summary of the Black population's distribution in the central region of Washington, DC. 

The parameter $\alpha$ is the local scaling exponent which describes how the density of cells scales with the cell size ($\epsilon$). 

\noindent $\bullet$ \ \  Low $\alpha$ values indicate that the densities of the daughter boxes (size $\epsilon / 2$) are only slightly smaller than those of the parent box (size $\epsilon$). Since the total density of all four daughter boxes must equal that of the parent box, low $\alpha$ implies that some daughter boxes are empty. This results in a heterogeneous, gappy pattern resembling a \textit{Swiss cheese}. On the map in Figure~1B, low $\alpha$ values are found in locations west of the Potomac River.

\noindent $\bullet$ \ \  High $\alpha$ values suggest that the densities of the daughter boxes are significantly smaller than that of the parent box. Given that the total density must remain the same, high $\alpha$ implies that most daughter boxes are non-empty. This results in a homogeneous pattern with few, if any, gaps. On the map in Figure~1B, high $\alpha$ values are located in areas east of the Potomac River.

The parameter $f(q)$ represents the fractal dimension of the subset of boxes scaling with the exponent $\alpha (q)$, indicating how much of the pattern exhibits a particular scaling behavior.

\noindent $\bullet$ \ \  Low $f(\alpha)$ values suggest that the scaling behavior characterized by a given $\alpha$ is present in fewer locations within the pattern. From Figure~\ref{multifractalSpectrum}, it is evident that low-$f(\alpha)$ areas are mainly associated with low-$\alpha$ values. Thus, regions with homogeneously distributed residency of the focus population tend to be spatially limited (dense areas in Figure~1B).
    
\noindent $\bullet$ \ \   High $f(\alpha)$ values indicate that the scaling behavior characterized by a given $\alpha$ is prevalent over a significant portion of the pattern. Examination of Figure~\ref{multifractalSpectrum} reveals that high-$f(\alpha)$ areas are predominantly linked with higher $\alpha$ values. Therefore, regions with heterogeneously distributed residency of the focus population cover more extensive areas (dispersed areas in Figure~1B).

 \begin{figure*}[t]
    \centering
     \includegraphics[width=0.93\textwidth]{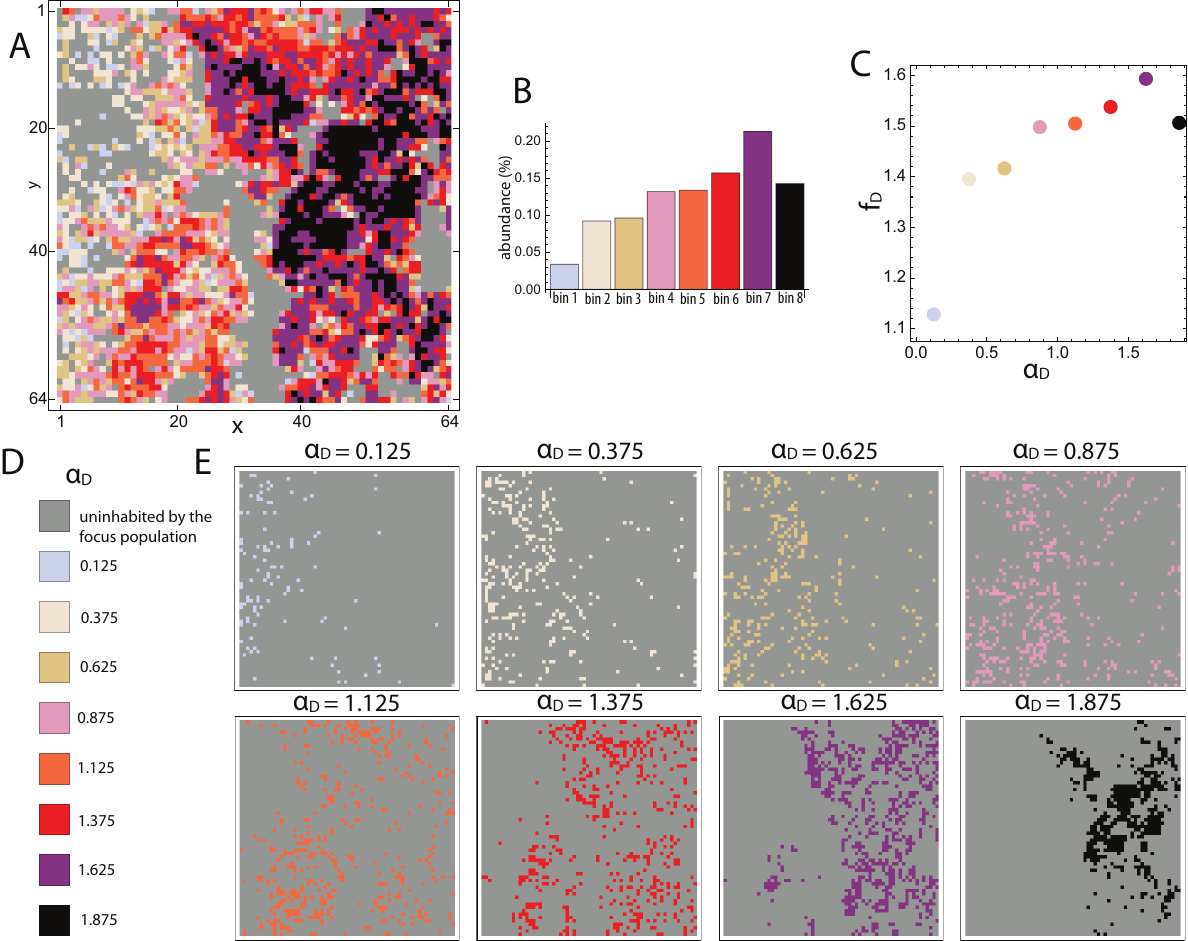}
    \caption{(A) Map of $\alpha_{\rm D}$ across the central region of Washington, DC. (B) Distribution of windows among sub-patterns based on different $\alpha_{\rm D}$ bins. (C) Discrete multifractal spectrum. (D) Legend for $\alpha_{\rm D}$ values. (E) Sub-patterns corresponding to different $\alpha_{\rm D}$ bins.
    \label{discreteMF}}
\end{figure*}

\subsection{Discrete multifractal spectrum.}
Because the spectrum described above is analytically derived with the value of $\alpha$ defined at a point, it cannot be used for mapping the heterogeneity of $\alpha$, and, ultimately, for mapping of heterogeneity of segregation, across the region \citep{Theiler1990}.

To enable mapping, we introduce the discrete multifractal spectrum. This variant allows for spatial visualization of multifractal properties. The procedure for constructing a discrete multifractal spectrum is as follows: First, the finest grid (\(\epsilon = 1/1024\)) of the region is partitioned into non-overlapping square windows, or local neighborhoods. The size of these neighborhoods must be small enough to maintain approximate homogeneity of cell densities within each window, yet large enough to encompass multiple smaller \(\epsilon\) grids, allowing for meaningful multifractal analysis. In this paper, we use windows corresponding to a grid size of \(\epsilon = 1/64\). This results in a grid of \(64 \times 64\) windows, each approximately \(0.5 \times 0.5\) km in size, containing \(16 \times 16 = 256\) original data cells.

Next, we calculate $\alpha_{\rm D}$ for each non-empty window; $\alpha_{\rm D}$ is the local density scaling exponent with \(\epsilon\). This exponent is derived from the scales available within the 16$\times$16 cells window, 1/64, 1/128, 1/256, 1/512, 1/1024. To determine values of $f_{\rm D}$, we classify the $\alpha_{\rm D}$ values into eight bins defined by the following breakpoints: $0.0, 0.25, 0.5, 0.75, 1.0, 1.25, 1.5, 1.75, 2.0$. This classification divides the pattern into eight non-overlapping sub-patterns, each grouping windows with similar $\alpha_{\rm D}$ values. The $f_{\rm D}$ values are computed as the fractal dimensions of these sub-patterns. The discrete multifractal spectrum is represented as a list of pairs \((\alpha_{\rm D}^j, f_{\rm D}^j)\), $j =1, \dots, 8$ , where $\alpha_{\rm D}^j$ is now the center value of bin $j$.

 \begin{figure*}[h]
	\centering
	\includegraphics[width=\textwidth]{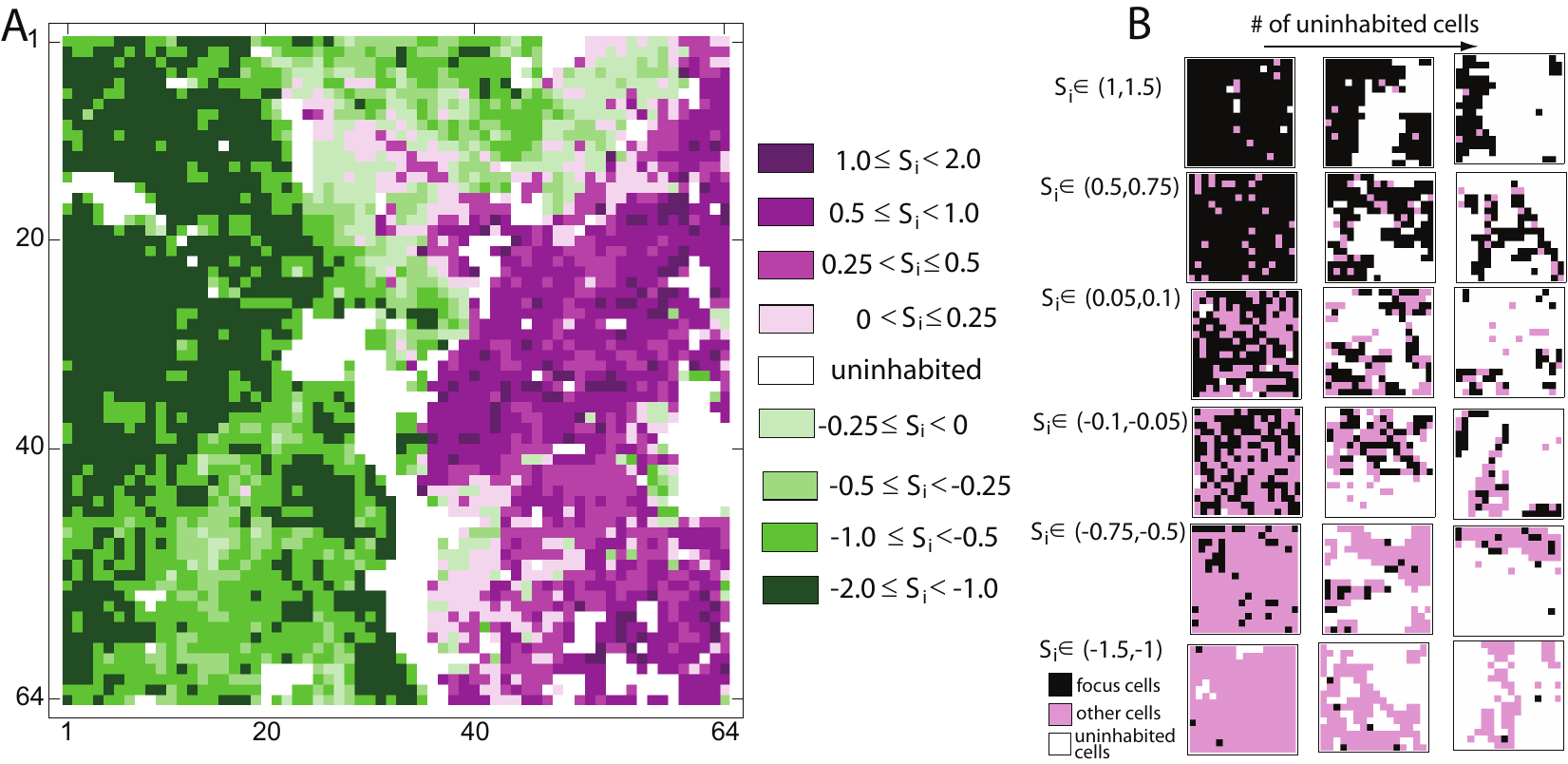}
	\caption{(A) Multifractal map of segregation (MFMS) for the Black sub-population in the central region of Washington, DC. This map illustrates segregation levels in local neighborhoods sized $0.5 \times 0.5$ km. Note the distinction between segregation as the exclusion of the non-focus population from a neighborhood (positive values of $S_i$) and segregation as the exclusion of the focus population from a neighborhood (negative values of $S_i$). (B) Examples of neighborhoods with similar $S_i$ values but varying numbers of uninhabited cells.
\label{segMap}}
\end{figure*}

Figure~\ref{discreteMF} presents the discrete multifractal spectrum for the Washington, DC region. Panel A illustrates the map of $\alpha_{\rm D}$, where the color of each window corresponds to the value of $\alpha_{\rm D}$. As colors transition from light to dark, the sub-patterns associated with them become progressively more homogeneous, exhibiting fewer gaps. Formally, the $\alpha_{\rm D}$ map shows the spatial variability of the gappiness of the Black residency pattern. From a demographic perspective, this map reveals the spatial variability in the interweaving of Black population residences with those of other populations and/or with uninhabited areas. This is because the gaps in the pattern are either filled with other populations or are uninhabited. Significant interweaving is indicated by light colors, while minimal interweaving is indicated by dark colors.

Panel B indicates the number of windows in each sub-pattern defined by $\alpha_{\rm D}$. Panel C presents the discrete multifractal spectrum, \((\alpha_{\rm D}^j, f_{\rm D}^j)\), $j =1, \dots, 8$. Note that the discrete multifractal spectrum differs from the analytic multifractal spectrum depicted in Fig.~2. This distinction arises because $\alpha$ in Fig.~2 is the density scaling exponent at a point, whereas $\alpha_{\rm D}$ represents the density scaling exponent within a $0.5 \times 0.5$ km window. Comparing panels B and C, we observe that the values of fractal dimension $f_{\rm D}$ (panel C) are proportional to the number of windows in each $\alpha_{\rm D}$-defined sub-pattern (panel B). Thus, we can use the value of $f_{\rm D}$ as a proxy for a relative size of sub-pattern associated with the value of $\alpha_{\rm D}$. Panel D provides the legend for $\alpha_{\rm D}$ values. Finally, Panel E separates the eight sub-patterns associated with different $\alpha_{\rm D}$ values for clarity.

\subsection{Multifractal Map of Segregation}
The final component of our proposed method of analysis is the Multifractal Map of Segregation (MFMS), which visualizes the spatial distribution of segregation within a city. As discussed in the previous subsection, the $\alpha_{\rm D}$ map illustrates the spatial variability in the interweaving of Black population residences with those of other populations and/or uninhabited areas. However, this depiction of segregation does not fully account for uninhabited areas, necessitating a separate segregation map.

To create a map showing within-city segregation variability, we apply a procedure similar to the construction of the discrete multifractal spectrum. We generate a $64 \times 64$ grid of windows. Instead of calculating the density scaling exponent for each window (as done for local values of $\alpha_{\rm D}$), we now compute the local fractal dimension for the focus population’s residency pattern ($D_{\rm fsp}$) and the local fractal dimension for the remaining population’s residency pattern ($D_{\rm rp}$). The key distinction between the density scaling exponent and the local fractal dimension is that the former pertains to the change in density with scale $\epsilon$, while the latter pertains to the change in the number of inhabited cells with $\epsilon$.

A measure of local segregation, $S_i$, within a window $i$ is defined as:
\begin{equation}
  S_i =  D_{\rm fsp}(i) - D_{\rm rp}(i)
\end{equation}

The MFMS displays the $S_i$ values of all windows across the region. Figure~\ref{segMap}A shows an example of an MFMS for the central region of Washington, DC. The range of $S_i$ values is from $-2$ to $2$. Positive $S_i$ values (shaded in violet) signify segregation characterized by the exclusion of populations other than the focus (Black) population. Conversely, negative $S_i$ values (shaded in green) signify segregation characterized by the exclusion of the focus (Black) population.

It is important to note that windows with similar $S_i$ values exhibit the same level of segregation, yet they may vary in the number of uninhabited boxes within the neighborhood. This is illustrated in Figure~\ref{segMap}B, which has six rows, each corresponding to a characteristic $S_i$ value. Each row presents three examples of windows (neighborhoods) that share similar $S_i$ values but differ in the number of uninhabited blocks. The segregation measure $S_i$ performs as expected. Its value is independent of the proportion of uninhabited blocks in the window; it relies solely on the ratio of focus to non-focus inhabited blocks and the gappiness of the pattern.

\begin{figure*}[h]
	\centering
	\includegraphics[width=\textwidth]{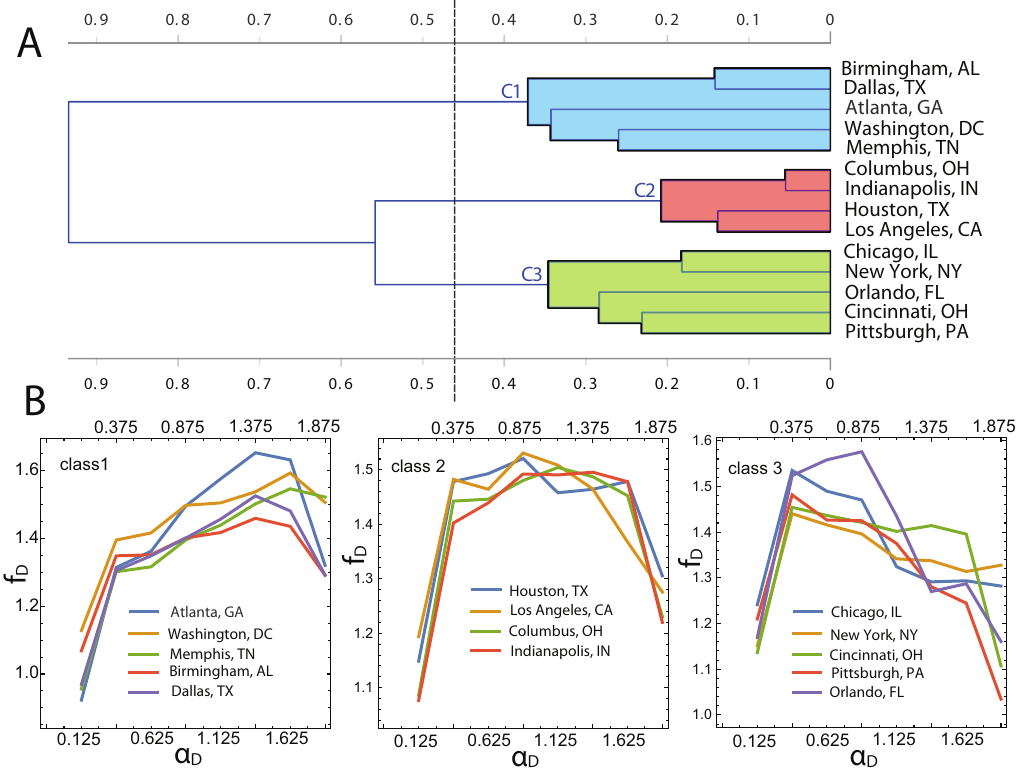}
	\caption{(A) Dendrogram resulting from clustering the discrete multifractal spectra for 14 residency patterns of the Black population in the central region of Washington, DC. Three classes of spectra, C1, C2, and C3, are identified by cutting the dendrogram at a selected dissimilarity level (indicated by the dotted line). (B) Discrete multifractal spectra for patterns classified into each class.
\label{dendro}}
\end{figure*}
\section{Results}
To illustrate the application of MFA for the quantitative and visual examination of residency patterns of the focus population, we analyze the residency patterns of the Black population across fourteen major U.S. cities (cities' names are listed in Figure~\ref{dendro}). Our goal is to explore the similarities in the residency patterns of the Black population and in their segregation from the rest of the population.

The dataset used for our analysis is derived from the National Racial Geography Dataset (NRGD2020) (available at \url{https://www.socscape.edu.pl}), which provides 30-meter resolution racial landscape data for the contiguous United States as of 2020. Specifically, we utilized two layers from the NRGD2020: a grid of racial IDs, where each cell is labeled with one of six racial categories, and a population density grid, where each cell indicates local population density. Notably, if a cell is labeled with a specific race in the racial ID grid, its population density in the corresponding cell of the second grid reflects the density of that race.

 \begin{figure*}[h]
	\centering
	\includegraphics[width=\textwidth]{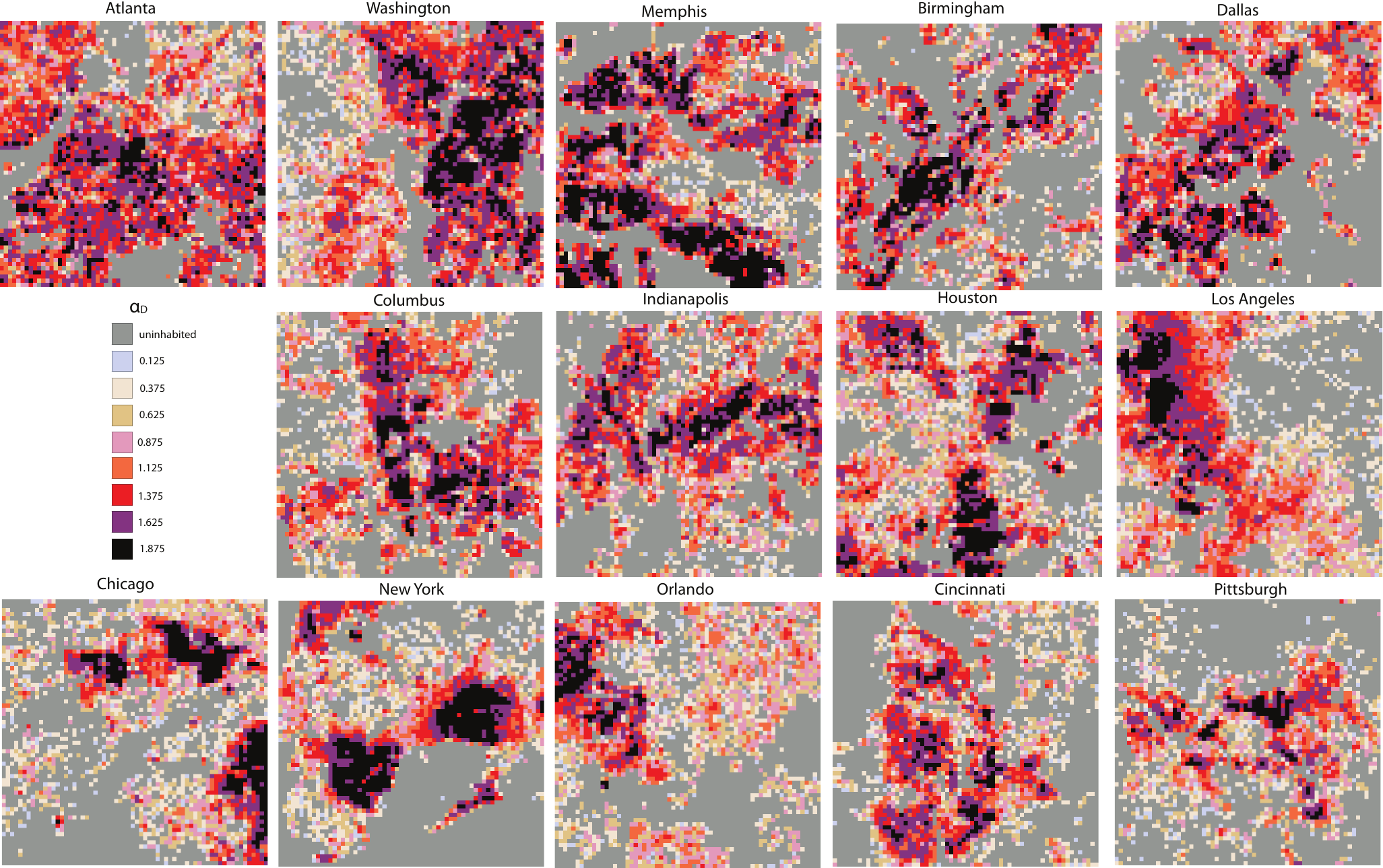}
	\caption{Maps showing heterogeneity of texture in 14 residency patterns of the Black population. Rows, from the top to the bottom, correspond to classes C1, C2, and C3.
\label{alphaMaps}}
\end{figure*}

For each city, we extracted a centrally located region of 1024$\times$1024 cells from the NRGD2020 dataset. These regions were converted into a ternary raster representing the focus population (1), the remaining population (-1), and uninhabited areas (0). Additionally, we created a raster of population density, specifying the density of the focus sub-population in cells labeled 1, and the density of the remaining population in cells labeled -1. Using the transformed data, we calculated the discrete multifractal spectra for the Black population in each city, following the method outlined in Section 2.1, and constructed multifractal segregation maps as described in Section 2.2.

We began by clustering the discrete multifractal spectra of the Black residency patterns using a hierarchical clustering algorithm. The Euclidean distance was employed to measure dissimilarity between discrete multifractal spectra, and the Ward linkage method \citep{ward1963} was used to gauge dissimilarity between clusters of spectra. Patterns are considered similar if their corresponding $\alpha_{\rm D}$ sub-patterns exhibit comparable $f_{\rm D}$ values, indicating similar structures of interweaving between the Black population's residences and those of other populations.

Figure~\ref{dendro}A presents a dendrogram summarizing the clustering results. This dendrogram, a tree-like diagram, displays the arrangement of clusters formed through hierarchical clustering. It consists of leaves and branches; each leaf represents an individual pattern (cities' names are shown on the right side of the dendrogram), and branches connect these leaves, indicating their inclusion in a cluster. Moving from right to left, branches merge, signifying the amalgamation of smaller clusters into larger ones. The horizontal axis reflects the dissimilarity between clusters, with shorter lengths indicating greater similarity and longer lengths indicating greater dissimilarity.

The dendrogram reveals that the branch lengths between individual leaves are relatively long, suggesting a lack of highly similar patterns. An exception is observed in the Black residency patterns of Columbus and Indianapolis, which, with a dissimilarity value of approximately 0.05, are notably more similar compared to other pairs such as Houston and Los Angeles or Dallas and Birmingham. Despite the general lack of high similarity among patterns, the dendrogram allows us to identify three distinct classes of patterns, designated as C1, C2, and C3.

\begin{figure*}[h]
	\centering
	\includegraphics[width=\textwidth]{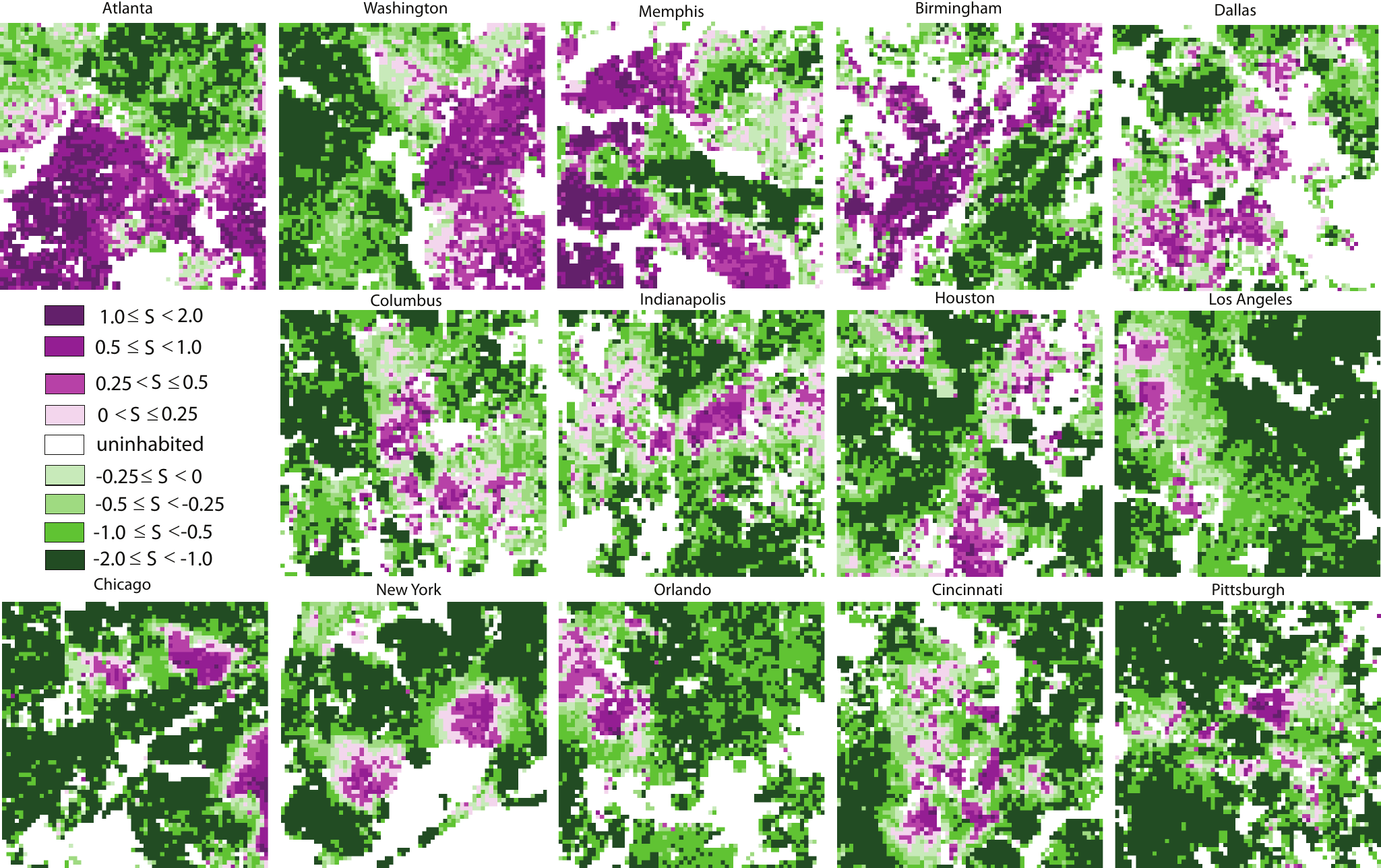}
	\caption{Maps showing heterogeneity of segregation in 14 residency patterns of the Black population. Rows, from the top to the bottom, correspond to classes C1, C2, and C3.
\label{examplesSegMaps}}
\end{figure*}

The spectra for each class are displayed in Figure~\ref{dendro}B. Across all classes, the values of $f_{\rm D}$ are relatively small at extreme values of $\alpha_{\rm D}$, reflecting the rarity of windows with extreme large or small gappiness. The distinguishing feature among the classes is how the abundance of windows ($f_{\rm D}$) changes with increasing $\alpha_{\rm D}$ outside of extreme values of $\alpha_{\rm D}$.

\noindent $\bullet$ \ \ In Class C1, $f_{\rm D}^j$ (proportional to the abundance of windows belonging to a bin $\alpha_{\rm D}^j$) is an increasing function of 
$\alpha_{\rm D}^j$. Cities classified as C1 have large swatches of land consisting of highly homogeneous Black neighborhoods. Neighborhoods with small presence of Black boxes are less abundant (see the top row in Figure~\ref{alphaMaps}). 

\noindent $\bullet$ \ \ In Class C3, $f_{\rm D}^j$ is a decreasing function of  $\alpha_{\rm D}^j$. Cities classified as C3 have large swatches of land consisting of only sporadic Black boxes. Neighborhoods with highly homogeneous Black neighborhoods are less abundant (see the bottom row in Figure~\ref{alphaMaps}).

\noindent $\bullet$ \ \ Class C2 shows a roughly constant abundance of windows for intermediate values of $\alpha_{\rm D}$, suggesting a balanced distribution of different levels of gappiness within the Black residency pattern (see the middle row in Figure~\ref{alphaMaps}).

Figure~\ref{examplesSegMaps}, organized similarly to Figure~\ref{alphaMaps}, presents maps of segregation of Black residences from the residencies of remaining population. While segregation maps are akin to $\alpha_{\rm D}$ maps, they provide additional details. Recall that $\alpha_{\rm D}$ is based solely on the focus (Black) residency pattern, whereas segregation maps also incorporate the residency patterns of the remaining population (as discussed in Section 2.2). 

The lightest colors on the segregation maps in Figure~\ref{examplesSegMaps}, those corresponding to $0 < S_i \leq 0.25$ and $-0.25 \leq S_i < 0$, represent mixed neighborhoods where the number of Black and other residents are roughly equal. In contrast, darker colors indicate more unbalanced populations, with neighborhoods dominated by either Black residences (positive $S_i$) or other populations (negative $S_i$).

In Figure~\ref{examplesSegMaps}, darker-colored neighborhoods are far more common than lighter-colored ones, indicating that the centers of all sampled cities are segregated. The three classes defined by the similarity of Black residency spectra (rows in Figure~\ref{examplesSegMaps}) show distinct spatial organizations of segregation. In Class C1 cities, Black residences are concentrated in a few large clusters, resulting in extensive zones of predominantly Black residences. Conversely, Class C3 cities exhibit many small clusters of Black residences. The segregation structures in Class C2 cities are more akin to those in Class C3 than to those in Class C1, as inferred from the dendrogram in Figure~\ref{dendro}A. 

\section{Discussion and conclusions}
In this paper, we have demonstrated how MFA—a well-established technique for statistical analysis and quantification of complex patterns—can be adapted to quantify residency patterns of racially-constrained populations and measure local segregation.

This work introduces several novel contributions:
\vspace{1mm}

\noindent $\bullet$ \ \ We utilize gridded data with monoracial cells as input, enabling the application of MFA and other innovative methods in racial geography. Although the creation of such data was not the focus of this paper, its application to racially-constrained populations is a novel approach.

\noindent $\bullet$ \ \ Unlike most prior studies, which apply MFA to point data (see references in the Introduction), we have applied MFA to gridded data. This choice aligns with the format of available racial data. Consequently, we employed the box-counting method for fractal analysis instead of the sandbox method \citep{tel1989,lengyel2023roughness}, which is directly mappable. Using box-counting method required the introduction of the discrete multifractal spectrum to obtain maps of $\alpha_{\rm D}$  and maps of segregation.

\noindent $\bullet$ \ \ From a racial demography perspective, the quantification and mapping of within-city segregation represent a significant advancement. Most previous studies calculated segregation as a single index for the entire city (see referenced in the Introduction). Using MFA, we characterize the residency pattern of the focus population across the city with the function $f_{\rm D}(\alpha_{\rm D})$ rather than a simple index. Furthermore, we map heterogeneity of segregation within a city through a grid of local segregation values.

Here is a list of answers to the questions that may arise about application of the MFA to racial geography.
\vspace{1mm}

\noindent $\bullet$ \ \ Input Data Availability: The racial landscape is available for the entire conterminous United States for the 2020 census year (\url{https://www.socscape.edu.pl}). For MFA applications to racial data from the 1990, 2000, and 2010 census years, the racial landscape must be first calculated using the R package \texttt{raceland}, available from The Comprehensive R Archive Network. MFA as described in this paper is limited to the US due to data availability rather than methodological constraints.

\noindent $\bullet$ \ \ Scope: Because this is a methodological paper, it focuses on the method rather than its applications. Thus, we selected extents of input data best-suited for calculations, while racial demographers would likely chose other extents.
However, our method work with data of any extent, including counties' and Metropolitan Statistical Areas' boundaries.

\noindent $\bullet$ \ \ Applications: The presented method is designed for the quantification and mapping of segregation variability within a city and is not intended to compete with existing methods for quantifying racial segregation for an entire city. The method addresses binary segregation, demonstrated for a single racially-constrained population versus the remaining population. Other combinations are feasible, provided only two populations are involved. The quantification and mapping of segregation in the context of a multiracial population using racial landscape data were previously discussed in \citep{Dmowska2024}.

\noindent $\bullet$ \ \ Interpretation of Figures~\ref{alphaMaps} and \ref{examplesSegMaps}: These figures are the final products of our method. It is important to interpret them correctly. The $\alpha_{\rm D}$ map may be mistaken for the map of density of Black (focus) population, but it is a map of gappiness of the pattern formed by residences of Black population. Theoretically, it does not depend on the density, but in practice, there is a correlation between the value of $\alpha_{\rm D}$ and the density. Areas with homogeneous distribution of Black residences tend to be the areas of high density of Black population. The segregation map, interpreted at the level of an individual window communicate a relative abundance of Black and non-Black populations. Thus, the value pf segregation does not depend on the density of Black population. Interpreted at the level of the entire region, the map shows the tendency of the two population not to mix with each other.

Lastly, the method described in this paper can be applied to spatial datasets other than population. Any categorical grid data can serve as input for our method. For example, other census data types can be analyzed using MFA after conversion to a monoracial grid. A prime non-urban candidate for MFA quantification is land cover data, which does not require conversion as it is already in a grid with mono-valued (categorical) cells. For example, MFA can be applied to study changes in tree cover patterns with climate, similar to \citep{Scanlon2007}, who investigated tree cover patterns in the Kalahari Desert at different sites along a rainfall gradient, or to temporal changes in land cover categories.

Data for the fourteen patterns used in the paper as well as a code enabling repeating the results shown in the paper are in the Supplementary file.



\end{document}